\PassOptionsToPackage{utf8}{inputenc}
\documentclass{bioinfo}
\usepackage{color}
\usepackage{multirow}
\usepackage{upgreek}
\usepackage{float}

\usepackage{multirow}
\usepackage{booktabs}
\usepackage[table]{xcolor}

\copyrightyear{2015} \pubyear{2015}

\access{Advance Access Publication Date: Day Month Year}
\appnotes{Manuscript Category}

\begin{document}
\firstpage{1}
\subtitle{Subject Section}
\title[SeekRBP]{SeekRBP: Leveraging Sequence-Structure Integration with Reinforcement Learning for Receptor-Binding Protein Identification}
\author[X. Luo and J. Shang \textit{et~al}.]{Xiling Luo$^{123}$, Le Ou-Yang$^{4}$, Yang Shen$^{2}$, Jiaojiao Guan$^{2}$, Dehan Cai$^{2}$, Jun Zhang$^{1}$, Guoliang Xing$^{5}$, Yanni Sun$^{2*}$, and Jiayu Shang$^{5*}$}
\address{
$^{\text{\sf 1}}$Guangdong Provincial Key Laboratory of Wastewater Information Analysis and Early Warning, Beijing Normal University, Zhuhai, China;\\
$^{\text{\sf 2}}$Department of Electrical Engineering, City University of Hong Kong, Hong Kong (SAR), China;\\
$^{\text{\sf 3}}$MoE Key Lab of Artificial Intelligence, AI Institute, School of Computer Science, Shanghai Jiao Tong University, Shanghai, China;\\
$^{\text{\sf 4}}$SMBU-MSU-BIT Joint Laboratory on Bioinformatics and Engineering Biology, Shenzhen MSU-BIT University, Shenzhen, China;\\
$^{\text{\sf 5}}$Department of Information Engineering, Chinese University of Hong Kong, Hong Kong (SAR), China}
\corresp{$^\ast$To whom correspondence should be addressed.}
\history{}
\editor{}
\vspace{-0.5cm}
\abstract{\textbf{Motivation:} Receptor-binding proteins (RBPs) initiate viral infection and determine host specificity, serving as key targets for phage engineering and therapy. However, the identification of RBPs is complicated by their extreme sequence divergence, which often renders traditional homology-based alignment methods ineffective. While machine learning offers a promising alternative, such approaches struggle with severe class imbalance and the difficulty of selecting informative negative samples from heterogeneous tail proteins. Existing methods often fail to balance learning from these ``hard negatives'' while maintaining generalization.\\
\textbf{Results:} We present SeekRBP, a sequence--structure framework that models negative sampling as a sequential decision-making problem. By employing a multi-armed bandit strategy, SeekRBP dynamically prioritizes informative non-RBP sequences based on real-time training feedback, complemented by a multimodal fusion of protein language and structural embeddings. Benchmarking demonstrates that SeekRBP consistently outperforms static sampling strategies. Furthermore, a case study on \textit{Vibrio} phages validates that SeekRBP effectively identifies RBPs to improve host prediction, highlighting its potential for large-scale annotation and synthetic biology applications. \\
\textbf{Availability:} The source code of SeekRBP is available via: https://github.com/Saillxl/SeekRBP.\\
\textbf{Contact:} \href{yannisun@cityu.edu.hk}{yannisun@cityu.edu.hk}, \href{jiayushang@cuhk.edu.hk}{jiayushang@cuhk.edu.hk}}
\maketitle

\section{Introduction}

\begin{figure*}[htbp]
    \centering
    \includegraphics[width=0.85\linewidth]{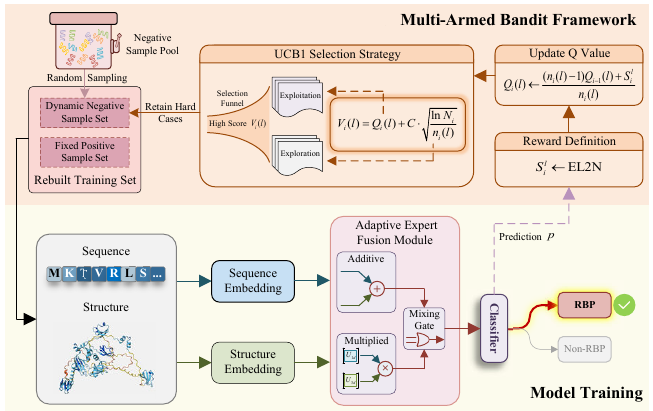} 
    \vspace{-0.5cm}
    \caption{Overview of the SeekRBP's dynamic negative sampling pipeline. A UCB1-based multi-armed bandit strategy iteratively selects informative hard negatives from a dynamic negative sample pool by balancing exploitation and exploration. At each round, the selected negatives are combined with the fixed positive set to reconstruct the training set. After model training, EL2N rewards are used to update their Q-values and guide the next round of selection.} Sequence and structure representations are adaptively fused and fed into a classifier to predict RBP versus non-RBP.
    \vspace{-0.5cm}
    \label{fig:framework}
\end{figure*}

Bacteriophages (or phages) represent the most abundant and genetically diverse biological entities on Earth \citep{suttle2007marine,mushegian2020there}. By infecting and regulating bacterial populations, they play a pivotal role in shaping microbial community structures and driving microbial evolution across diverse ecosystems \citep{sutton2019gut}. Many ecological and applied functions of phages, including phage therapy, biocontrol, and microbiome modulation, are mediated primarily through specific interactions with their bacterial hosts. These interactions are encoded within phage genomes, which harbor a vast repertoire of proteins with antimicrobial and biotechnological potential. However, the majority of phage-encoded proteins remain functionally uncharacterized due to limited and uneven annotation resources \citep{lu2021prokaryotic,nayfach2021checkv}. This persistent annotation gap constitutes a significant bottleneck for the effective utilization of phages in therapeutic, industrial, and biocontrol applications, particularly given the escalating global burden of antimicrobial resistance \citep{gordillo2019phage,murray2022global}.

Among phage-encoded proteins, receptor-binding proteins (RBPs) are key determinants of infection specificity and host range \citep{dowah2018review}. Typically localized at the distal tips of tail fibers or tail spikes, RBPs mediate the initial adsorption step by recognizing and binding to bacterial surface receptors \citep{nobrega2018targeting}. Although several RBP families are curated in databases such as pVOGs \citep{grazziotin2016prokaryotic} and PHROGs \citep{terzian2021phrog}, existing annotations remain sparse and disproportionately biased toward well-studied model phage systems. This limitation is exacerbated by the structural heterogeneity of RBPs, which vary significantly in length, domain composition, and sequence architecture; many also contain repetitive or low-complexity regions that complicate computational discovery \citep{boeckaerts2021predicting}. A primary challenge is the high sequence diversity and rapid evolutionary turnover of RBPs, driven by phage--host co-evolutionary dynamics \citep{hampton2020arms,scanlan2015coevolution}. Even RBPs targeting the same bacterial species often exhibit negligible sequence similarity, limiting the efficacy of sequence similarity- or profile-based annotation methods in identifying remote or novel RBPs \citep{pearson2013introduction}.

\vspace{-0.3cm}
\subsection{Existing work}
Several attempts have been made to identify RBPs within phage genomes. Traditional annotation relies on homology-based alignment tools, such as BLASTp \citep{altschul1990basic}, and HMM profiles (e.g., Pharokka \citep{bouras2023pharokka}), which search against reference databases. While effective for conserved sequences, these methods are limited by the rapid evolutionary turnover and high divergence characteristic of phage RBPs, often failing to detect remote homologs. Deep learning approaches, including PhANNs \citep{cantu2020phanns} and PhageRBPdetection \citep{boeckaerts2022identification}, have been developed to extract latent patterns from raw sequences, offering improved generalization over profile-based methods. Furthermore, DeepFRI \citep{gligorijevic2021structure} has successfully employed Graph Convolutional Networks (GCNs) to leverage structural information for function prediction. More recently, Transformer-based architectures such as DePP \citep{magill2023depolymerase} have utilized self-attention mechanisms to capture long-range dependencies in viral proteomes. However, the performance of these models is often constrained by significant class imbalance. In typical phage genomes, RBPs constitute only a small fraction of coding sequences. Consequently, predictive models tend to be biased toward the majority class, resulting in poor recall for true RBPs in practical applications \citep{chicco2020advantages}.

\subsection{Overview}
In this work, we present SeekRBP, a deep learning framework designed to identify phage RBPs under conditions of extreme class imbalance. Specifically, SeekRBP reformulates the identification task as a sequential decision-making process rather than a static classification problem. To address the issues where current models are biased by majority-class signals or limited by high sequence divergence, we made two major contributions. First, we introduce an adaptive sampling strategy inspired by Reinforcement Learning (RL) to dynamically select informative negative samples. To effectively handle the overwhelming number of non-RBP proteins (negative samples), SeekRBP treats training as an adaptive process; the framework dynamically identifies and prioritizes ``informative/hard'' negative samples, specifically, non-RBPs that share sequence or structural similarity with true RBPs. By continuously exposing the model to these challenging examples, we prevent bias toward the majority class and compel the network to learn robust, more discriminative decision boundaries.

Second, we developed a dual-branch architecture that integrates both sequence- and structure-derived representations to capture complementary functional signals. Because phage-host co-evolution drives rapid sequence mutation, RBPs often exhibit highly divergent amino acid sequences despite performing identical functions. However, 3D structural features can provide supplementary information that enhances RBP identification. To leverage this, SeekRBP utilizes a specialized module that fuses 1D sequence data with 3D structural features. This integration enables the framework to recognize the physical determinants of receptor binding, such as the structural conformation of the tail tip, even in remote homologs where the primary sequence has diverged substantially.

We validated SeekRBP through comprehensive experiments on rigorously designed benchmark datasets. The results demonstrate that our framework consistently yields higher sensitivity and accuracy compared to state-of-the-art baselines. Notably, by combining adaptive learning with structural insights, SeekRBP successfully identifies novel RBPs that are typically overlooked by existing methods and enhance the performance of host prediction.

\section{Materials and methods}

We formulate the identification of RBPs as a RL task to address the fundamental challenge of learning under extreme class imbalance. This paradigm treats the selection of negative samples not as a static step, but as a dynamic process where the utility of training data evolves during optimization. Unlike standard supervised learning, which typically assumes a fixed dataset with uniform informativeness, our approach focuses on selecting ``hard'' negative samples relative to the current model state. This strategy is particularly well-suited to RBP identification, where the vast majority of genomic sequences are non-RBPs that provide little gradient signal. Thus, by adaptively prioritizing ``informative'' negatives based on real-time feedback, the model can effectively filter out background noise and focus on the challenging instances that define the true decision boundary.

In the following sections, we first describe the overall SeekRBP framework (Fig. \ref{fig:framework}), detailing the adaptive negative sampling module modeled as a multi-armed bandit problem. Next, we introduce the dual-branch representation architecture, explaining how complementary features are extracted from both protein sequences and predicted 3D structures. Finally, we outline the fusion and classification module designed to integrate these multimodal signals for accurate RBP identification.

\subsection{Bandit-based adaptive negative sampling framework}
To facilitate efficient adaptation during training, we formulate the selection of negative samples as a multi-armed bandit (MAB) problem. In this setting, each candidate negative sample is treated as a distinct ``arm'' (or option). The utility of each candidate is estimated based on feedback received during training and is updated iteratively. This process balances the \textit{exploitation} of known informative negatives (often examples sharing sequence or structure similarity with RBPs) with the \textit{exploration} of under-sampled (unseen) candidates. 

In our framework (Fig. \ref{fig:framework}), negative sample selection is modeled as a sequential interaction between a sampling agent and the training process. At each iteration, the agent selects a subset of negative samples for model updates, and the utility of these samples is assessed based on feedback from the model. Negatives that prove informative are prioritized in subsequent iterations, while redundant or easy samples are progressively down-weighted. This design allows the sampling policy to co-evolve with the model, adapting dynamically rather than remaining fixed throughout training.

\subsubsection{Dataset construction}
We constructed the phage protein dataset using sequences from the INPHARED phage genome database \citep{cook2021infrastructure}. The initial corpus contained 2,484,471 phage protein sequences annotated with PHROGS \citep{terzian2021phrog}. To ensure high annotation quality, we removed sequences labeled as hypothetical, putative, unknown, or unannotated. We then de-duplicated the remaining sequences using CD-HIT \citep{fu2012cd} with a 98\% sequence identity threshold, yielding 266,311 non-redundant protein sequences. Then, we follow the definition of the RBP and label the sequences based on structural annotations. Proteins annotated as \textit{tail fiber} or \textit{tail spike} were designated as positive samples. To construct a conservative negative set and minimize label noise, we excluded sequences broadly annotated as \textit{tail} that lacked specific \textit{fiber} or \textit{spike} descriptors. All remaining non-RBP sequences were treated as negative samples. The final dataset comprises 11,096 positive and 212,642 negative samples, resulting in a class imbalance of approximately 5\% positives. 

To ensure a rigorous evaluation and prevent data leakage arising from sequence homology, we clustered sequences separately within the positive and negative classes using CD-HIT at a 40\% sequence identity threshold. These clusters, rather than individual sequences, were partitioned into training, validation, and test sets (80\%-10\%-10\% split), ensuring that homologous sequences appear in only one data split.

\subsubsection{Bandit-based negative sample selection and utility estimation}
Our adaptive negative sampling strategy employs the Upper Confidence Bound (UCB1) algorithm \citep{auer2000using}. This approach provides a principled mechanism to balance the \textit{exploitation} of known informative samples against the \textit{exploration} of under-sampled ones. In this framework, each candidate negative sample $l$ is treated as a distinct ``arm'', with its utility estimated and updated throughout the training process. UCB1 combines each sample's historical utility with an uncertainty bonus for rarely selected samples. High-scoring negatives are selected for training, and their utilities are updated using EL2N (Error L2 Norm) \citep{paul2021deep} rewards. Positive samples are excluded from this bandit selection; they are always included in the training batch to ensure stable supervision given their scarcity.

\begin{figure*}[htbp]
    \centering
    \includegraphics[width=0.95\linewidth]{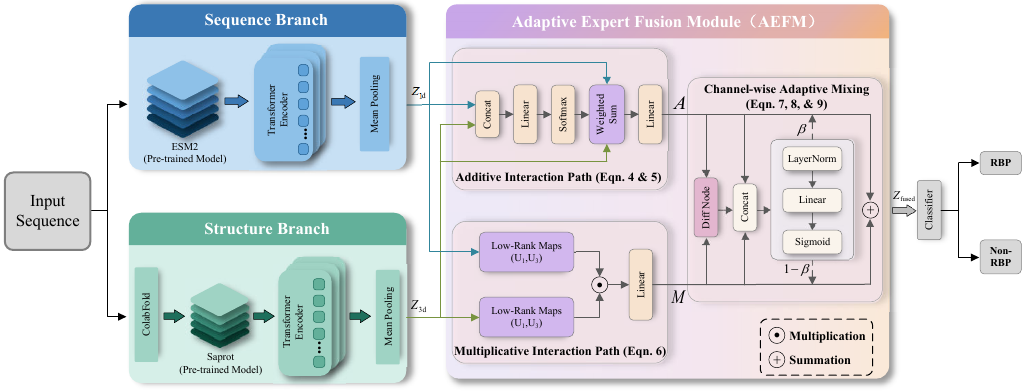}
    \vspace{-0.5cm}
    \caption{Pipeline of the proposed model for phage RBP identification. The input protein is encoded by a sequence branch and a structure branch, whose features are integrated by an Adaptive Expert Fusion Module combining gated additive and low-rank bilinear interactions. The fused representation is fed into a classifier to predict RBP versus non-RBP.}
    \label{fig:model}
    \vspace{-0.5cm}
\end{figure*}

At each iteration $i$, we compute a selection score $V_i(l)$ for each candidate negative sample as listed in Eqn. \ref{eqn:1}.

\vspace{-0.3cm}
\begin{equation}
\label{eqn:1}
    V_i(l) = Q_i(l) + C \cdot \sqrt{\frac{\ln N_i}{n_i(l)}}
\end{equation}
\vspace{-0.3cm}

\noindent where $Q_i(l)$ represents the estimated utility of negative sample $l$, $n_i(l)$ is the number of times sample $l$ has been selected up to iteration $i$, and $N_i$ is the total number of selections made across all samples. The hyperparameter $C$ controls the trade-off between exploitation and exploration.

The score $V_i(l)$ integrates two complementary objectives. The first term, $Q_i(l)$, quantifies the historical utility of sample $l$. In practice, negatives that are frequently misclassified or yield high error gradients receive larger $Q_i(l)$ values; these act as ``hard negatives'' that are critical for refining the decision boundary. The second term, derived from the count $n_i(l)$, accounts for uncertainty. It promotes exploration by inflating the scores of rarely selected samples, thereby preventing the model from overfitting to a narrow subset of hard cases and ensuring broad coverage of the negative space.

To estimate $Q_i(l)$, we maintain a running average of the utility based on feedback from the training dynamics. Following each training update at round $i$, we compute an immediate reward from the current model prediction using the EL2N metric (Eqn.~\ref{eqn:2}).

\vspace{-0.3cm}
\begin{equation}
\label{eqn:2}
S_i(l) =
\sqrt{
\sum_{c=1}^{K}
\left(
p_c(w_i,l)-y_{l,c}
\right)^2
}
\end{equation}
\vspace{-0.3cm}

\noindent
where $p_c(w_i,l)$ denotes the predicted probability of 
sample $l$ belonging to class $c$ under the model parameters $w_i$, 
$y_{l,c}$ is the corresponding element of its label, and $K$ denotes the number of output classes. A larger $S_i(l)$ indicates a significant discrepancy between the prediction and the label, suggesting that the sample is currently informative for the model.

Then, the utility estimate $Q_i(l)$ is updated iteratively (Eqn. \ref{eqn:3}). This rule provides a stable, long-term estimate of sample utility, allowing the bandit policy to adapt dynamically to the model's evolving needs. In each round, a batch of high-scoring negative samples is selected based on $V_i(l)$, ensuring that the model continuously focuses on the most instructive data points.

\vspace{-0.3cm}
\begin{equation}
\label{eqn:3}
    Q_i(l) \leftarrow \frac{(n_i(l) - 1)Q_{i-1}(l) + S_i(l)}{n_i(l)}
\end{equation}
\vspace{-0.3cm}

\subsection{Model architecture}

We propose a dual-branch model for the binary classification of RBPs in phages. As shown in Fig. \ref{fig:model}, the model comprises two parallel feature-encoding branches, a feature fusion module, and a classification module. The sequence branch encodes sequence-derived features, whereas the structure branch encodes structural and spatial features. Each branch produces a fixed-length, protein-level representation, which is then integrated by the fusion module to learn cross-modal information. The fused representation is finally fed into a multilayer perceptron (MLP) classifier to generate predictions.

\subsubsection{Dual-branch protein encoding}

\paragraph{The 1D sequence branch}
The sequence branch uses ESM2 \citep{lin2023evolutionary}, a pre-trained protein language model, to extract representations from input protein sequences. Given an input sequence, ESM2 produces a high-dimensional embedding for each residue position. The embeddings are then processed by a Transformer encoder to obtain 1D sequence hidden representations $\mathbf{Z}_{1d}$ (see Supplementary S1.1).

\paragraph{The 3D structure branch}
The structure branch encodes structure-derived features obtained by first predicting three-dimensional structures with ColabFold \citep{mirdita2022colabfold} and then extracting structure-aware representations using Saprot \citep{su2023saprot}. Specifically, ColabFold generates a predicted 3D structure for each input sequence, which is then processed by Saprot to produce residue-level structural features. These features are fed into a
Transformer encoder to obtain 3D structure representations $\mathbf{Z}_{3d}$ (see Supplementary S1.2).

\subsubsection{Adaptive expert fusion module (AEFM)}

The proposed module provides an adaptive trade-off between the stability of additive feature fusion \citep{he2016deep} and the stronger expressive capacity of multiplicative interaction modeling \citep{jayakumar2020multiplicative, perez2018film}, thereby enabling effective integration of sequence-derived and structure-derived representations. Inspired by the \emph{Mixture-of-Experts} (MoE) paradigm \citep{shazeer2017outrageously}, additive and multiplicative interaction forms are treated as parallel experts, and lightweight gating mechanisms dynamically control their contributions. This design improves cross-modal representation learning while preserving a compact and efficient model structure.

The fusion stage jointly considers two complementary interaction paradigms, namely \emph{additive interaction} and \emph{multiplicative interaction}. An adaptive gating mechanism dynamically balances these two forms to combine stability with expressive capacity.

\paragraph{Additive Interaction Path:}
The representations from the sequence branch and the structure branch are concatenated as $\mathbf{x} \in \mathbb{R}^{B \times 2d}$, where $d$ denotes the feature dimension. A gating network predicts the relative importance of the two feature streams, followed by softmax normalization:
\begin{equation}
[\alpha_{1d}, \alpha_{3d}] = \mathrm{softmax}(\varphi_1(\mathbf{x})) \in \mathbb{R}^{B \times 2},
\end{equation}
where $\varphi_1(\cdot)$ denotes a linear layer. The two feature representations are combined through a weighted sum and projected into the fusion space:
\begin{equation}
\mathbf{A} = \varphi_2(\alpha_{1d} \odot \mathbf{Z}_{1d} + \alpha_{3d} \odot \mathbf{Z}_{3d}) \in \mathbb{R}^{B \times d},
\end{equation}
where $\odot$ denotes element-wise multiplication, and $\varphi_2(\cdot)$ denotes a projection pathway composed of Layer Normalization, Dropout, and a Linear layer.

\paragraph{Multiplicative Interaction Path:}
To capture higher-order dependencies between sequence-derived and structure-derived features, we introduce a low-rank multiplicative interaction mechanism. The two feature streams are first projected into a low-rank space:
\begin{equation}
\mathbf{M} = \varphi_3 ((\mathbf{Z}_{1d} U_{1d}) \odot (\mathbf{Z}_{3d} U_{3d})) \in \mathbb{R}^{B \times d},
\end{equation}
where $\mathbf{U}_{1d}, \mathbf{U}_{3d} \in \mathbb{R}^{d \times r}$ are learned weight matrices and $r$ denotes the rank dimension. The output $\mathbf{M} \in \mathbb{R}^{B \times d}$ is then obtained via a linear layer $\varphi_3(\cdot)$.

\begin{figure*}[htbp]
    \centering
    \includegraphics[width=1.0\linewidth]{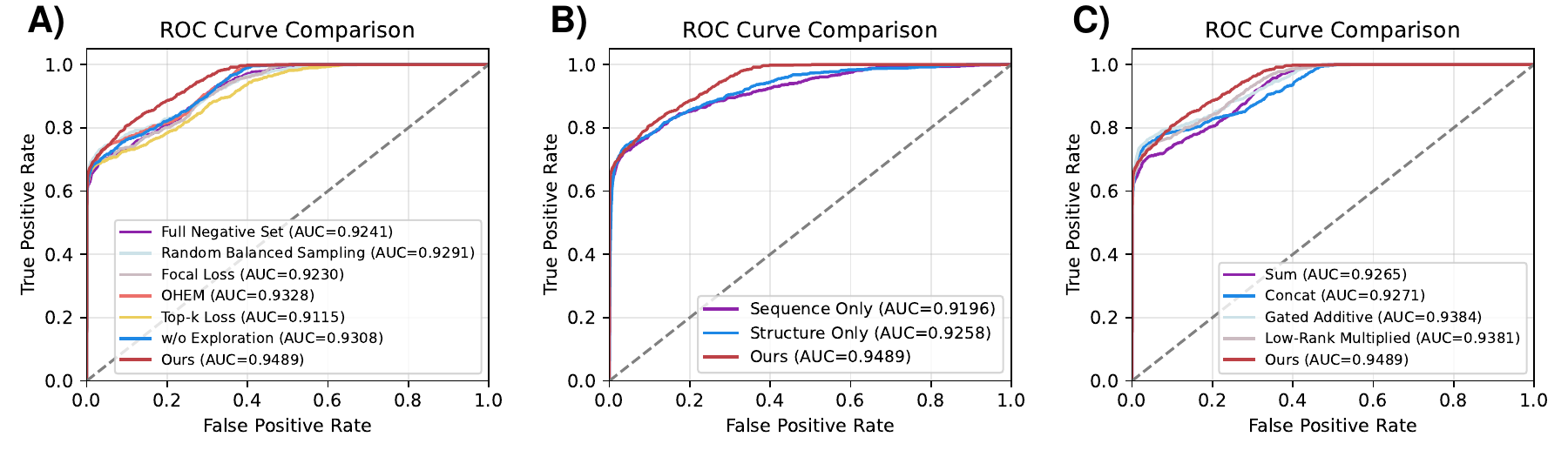}
    \vspace{-0.5cm}
    \caption{ROC curve comparison of ablation study. A) Sampling strategies. The \textit{w/o Exploration} variant represents a degenerate form of the bandit-based strategy.  B) Different feature modalities. \textit{Sequence Only}: uses sequence-derived features; \textit{Structure Only}: uses structure-derived features; and sequence combined both modalities (denoted as Ours). C) Feature fusion strategies.  \textit{Sum}: additive fusion; \textit{Concat}: concatenation; \textit{Gated Additive}: only use additive interaction path; \textit{Low-Rank Multiplied}: only use multiplicative interaction path; \textit{Ours}: adaptive expert fusion module.}
    \vspace{-0.5cm}
    \label{fig:ablation_study}    
\end{figure*}

\paragraph{Channel-wise Adaptive Mixing:}
Because different samples---and different feature channels within a sample---can benefit from different fusion strategies, a channel-wise adaptive mixing gate is introduced. The gate takes as input the concatenation of additive features, multiplicative features, and their absolute difference:
\begin{equation}
\mathbf{u} = [\mathbf{A}; \mathbf{M}; |\mathbf{A} - \mathbf{M}|]
\in \mathbb{R}^{B \times 3d}.
\end{equation}
Here, $|\mathbf{A}-\mathbf{M}|$ provides a direction-independent, channel-wise measure of disagreement between the additive and multiplicative representations. Together with $\mathbf{A}$ and $\mathbf{M}$, this discrepancy signal enables the gate to adaptively determine their relative contributions for each feature channel. The gating function produces channel-wise mixing coefficients:
\begin{equation}
\boldsymbol{\beta} = \sigma(\varphi_4(\mathbf{u})) \in \mathbb{R}^{B \times d},
\end{equation}
where $\sigma(\cdot)$ denotes the sigmoid function and $\varphi_4(\cdot)$ denotes a linear projection with Layer Normalization. The gating coefficients $\boldsymbol{\beta}$ are learned during training and are computed in a sample-specific manner based on the input features. The final fused representation is obtained via channel-wise interpolation:
\begin{equation}
\mathbf{Z}_{\text{fused}} =
\boldsymbol{\beta} \odot \mathbf{A} +
(1 - \boldsymbol{\beta}) \odot \mathbf{M}
\in \mathbb{R}^{B \times d}.
\end{equation}
This mechanism allows the model to adaptively emphasize additive fusion (larger $\boldsymbol{\beta}$) or multiplicative interaction (smaller $\boldsymbol{\beta}$) depending on the input features. The fused representation $\mathbf{Z}_{\text{fused}}$ is then fed into a lightweight classifier for final prediction. 
The classifier consists of Layer Normalization, Dropout, and a two-layer MLP with GELU activation. 
It outputs logits corresponding to the RBP and non-RBP classes. 
The classifier is intentionally kept lightweight.

\subsection{Training strategy and evaluation metrics}

During the training process, the selected negative samples and all positive samples constitute a training batch used to update the model parameters. Following this update, the model's predictions are refreshed, and the reward signal, specifically the prediction discrepancy $S^l$ (Eqn. \ref{eqn:2}), is recomputed to reflect the current state of the model. The model is trained using the Adam optimizer with a learning rate of $10^{-3}$ for a maximum of $30$ epochs. This iterative loop continues until the performance gain plateaus or a maximum number of epochs is reached. At the end of each cycle, the utility estimates $Q_i(l)$ and selection counts $n_i(l)$ are updated, informing the sampling policy for the subsequent round. More details are presented in the Supplementary S1.3.

\paragraph{Metrics}
In binary classification, the Area Under the Receiver Operating Characteristic Curve (AUROC) serves as a widely accepted metric for evaluating model performance. We also plot the receiver operating characteristic (ROC) curve to illustrate the diagnostic ability of the classifiers by mapping its true positive rate (model’s ability to correctly identify RBPs) against the false positive rate (frequency of errors in which non-RBPs are misclassified as RBPs). Given the class imbalance in RBP identification, we additionally evaluate performance using precision--recall (PR) curves and average precision (AP), which emphasize the reliability of positive predictions. Detail explanation about the calculation can be found in Supplementary S1.4.

\section{Results}
In this section, we evaluated the performance of SeekRBP through a series of computational experiments. We first conducted ablation studies on the validation set to quantify the contributions of key components, including the bandit-based negative sampling strategy, the integration of sequence and structural features, and the proposed fusion mechanism. Subsequently, we benchmarked SeekRBP on the test set against several established tools for phage protein annotation and RBP identification: PhANNs, PhageRBPdetection, Pharokka, and BLASTp. To ensure a fair comparison, all baselines were retrained or customized on the same dataset and assessed using identical data splits and evaluation protocols. Finally, we presented case studies using independent experimental datasets of \textit{Vibrio} phages to demonstrate the practical utility of SeekRBP in downstream applications.

\subsection{Ablation study}
\subsubsection{Effect of sampling strategies}
First, we investigated the impact of different sampling strategies on model performance. Fig. \ref{fig:ablation_study} A indicates that our bandit-based sampling strategy achieves the highest AUROC, particularly in regions with low false positive rates. Specifically, it achieves an AUROC of 0.9489, outperforming OHEM (0.9328), Focal Loss (0.9230), Random Balanced Sampling (0.9291), and Top-$k$ Loss (0.9115). The \textit{w/o Exploration} variant represents a degenerate form of the bandit-based strategy where the exploration term in Eq. \ref{eqn:1} is removed. While this exploitation-only strategy improves upon uniform sampling, it is consistently outperformed by the full UCB-based approach, highlighting the importance of explicitly encouraging the exploration of under-sampled negatives. Furthermore, we observe that the disparity between sampling strategies is sensitive to the input modality (Fig. S4 in supplementary).

\subsubsection{Contribution of sequence and structure features}
Next, we evaluated the relative contributions of sequence and structure features. As illustrated in Fig. \ref{fig:ablation_study} B, the model utilizing only structure features achieves an AUROC of 0.9258, whereas the sequence-only model yields a slightly lower AUROC of 0.9196. This suggests that while sequence features are commonly regarded as informative, structure data may offer greater discriminative power in our identification task. Notably, the integration of both modalities results in the highest performance (AUROC 0.9489), consistently outperforming single-modality approaches. One plausible reason is that the model can utilize the complementary nature of the two features: sequence data provides broad discriminative signals, while structural features help resolve ambiguous cases that are difficult to distinguish via sequence alone. Thus, multimodal integration is essential for maximizing the robustness and accuracy of RBP classification.

\subsubsection{Effectiveness of feature fusion strategies}

Finally, we evaluated the effectiveness of different strategies for fusing sequence and structure representations. As shown in Fig. \ref{fig:ablation_study} C, simple feature concatenation (Concat) yields an AUROC of 0.9271, indicating that simply stacking features offers limited gains. Similarly, additive fusion (Sum) achieves an AUROC of 0.9265, suggesting that these naive aggregation methods remain insufficient for fully capturing cross-modal complementarity. In contrast, more advanced strategies consistently enhance performance: gated additive fusion increases the AUROC to 0.9384 by adaptively reweighting features, while low-rank multiplicative fusion achieves an AUROC of 0.9381, highlighting the benefits of modeling nonlinear cross-modal interactions. Ultimately, our proposed AEFM method achieves the highest AUROC of 0.9489. By dynamically combining additive and multiplicative interactions, AEFM integrates sequence and structural information more effectively than any single fusion strategy.

\subsection{Comparison with benchmark methods}
\begin{figure}[htbp]
    \centering
    \vspace{-0.5cm}
    \includegraphics[width=0.9\linewidth]{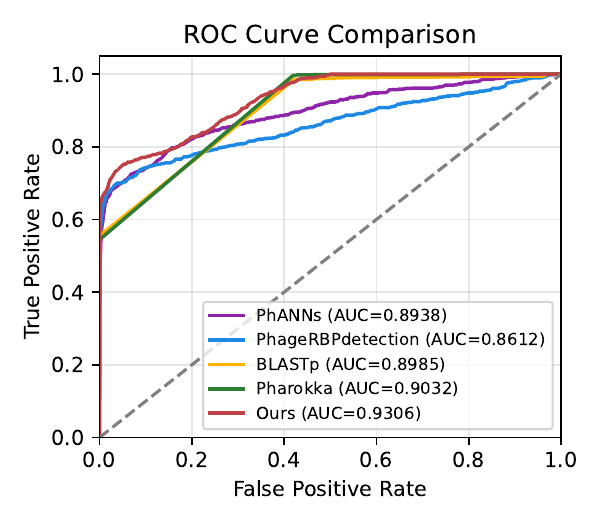} 
    \vspace{-0.7cm}
    \caption{ROC curves evaluating the predictive performance of SeekRBP (denoted as "Ours") alongside existing machine learning-based tools (PhANNs, PhageRBPdetection) and homology-based methods (BLASTp, Pharokka). The ROC curve of the homology-based method appears as a piecewise linear segment rather than a smooth curve, owing to the discrete and coarsely distributed prediction scores.}
    \vspace{-0.5cm}
    \label{fig:roc_sota}
\end{figure}

To further evaluate the predictive performance of SeekRBP, we benchmarked it against several state-of-the-art approaches, encompassing both machine learning-based tools (PhANNs and PhageRBPdetection) and homology-based methods (BLASTp and Pharokka). Following widely adopted practices, we utilized the $roc\_curve$ function within the scikit-learn library to construct our ROC curves: for the learning-based models, ROC curves were generated using the output prediction probabilities, whereas alignment bit-scores were utilized for the homology-based methods. As illustrated in Fig. \ref{fig:roc_sota}, SeekRBP achieves the highest AUROC of 0.9306, consistently outperforming all baseline methods. Notably, the homology-based methods, Pharokka and BLASTp, demonstrate strong performance with AUROC scores of 0.9032 and 0.8985, respectively, surpassing the existing learning-based tools. This performance suggests that our bandit-based sampling strategy effectively identifies highly informative negative examples, enabling the model to learn more discriminative features of RBPs without sacrificing generalization to divergent RBPs.

\begin{figure}[htbp]
    \centering
    \vspace{-0.5cm}
    \includegraphics[width=0.9\linewidth]{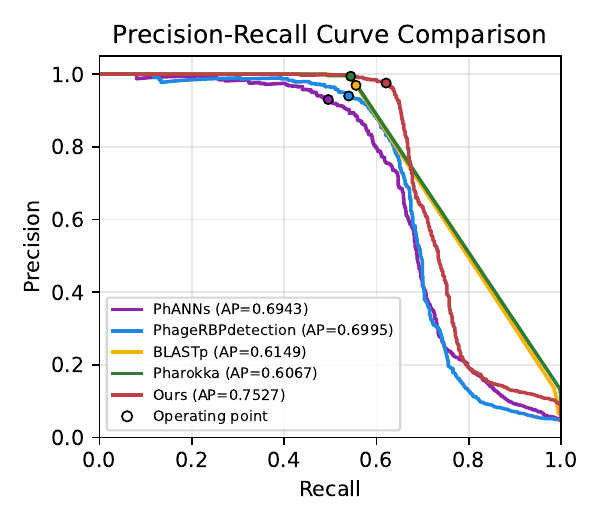}
    \vspace{-0.7cm}
    \caption{PR curves comparing the predictive performance of SeekRBP (denoted as ``Ours'') with existing machine learning-based tools (PhANNs and PhageRBPdetection) and homology-based methods (BLASTp and Pharokka). The AP for each method is reported in the legend, and the circular markers indicate performance at the default classification thresholds. The curves of the homology-based methods appear as piecewise linear segments owing to their discrete and coarsely distributed prediction scores. These connecting line segments are shown for visualization only.}
    \vspace{-0.5cm}
    \label{fig:pr_sota}
\end{figure}

Given the pronounced class imbalance in the RBP identification task, we additionally evaluated all methods using PR curves and AP (Fig. \ref{fig:pr_sota}). AP was computed without linear interpolation using the \texttt{average\_precision\_score} function in scikit-learn. SeekRBP achieved an AP of 0.7527, the highest among the evaluated methods, and maintained high precision across a substantial recall range. SeekRBP achieved higher recall at the default threshold while retaining precision comparable to the strongest baselines. This operating profile is particularly relevant to genome-wide RBP screening because RBPs constitute only a small fraction of phage proteins, meaning that even a modest reduction in precision can generate a large number of false-positive candidates. 

\begin{figure}[htbp]
    \centering
    \vspace{-0.5cm}
    \includegraphics[width=1.0\linewidth]{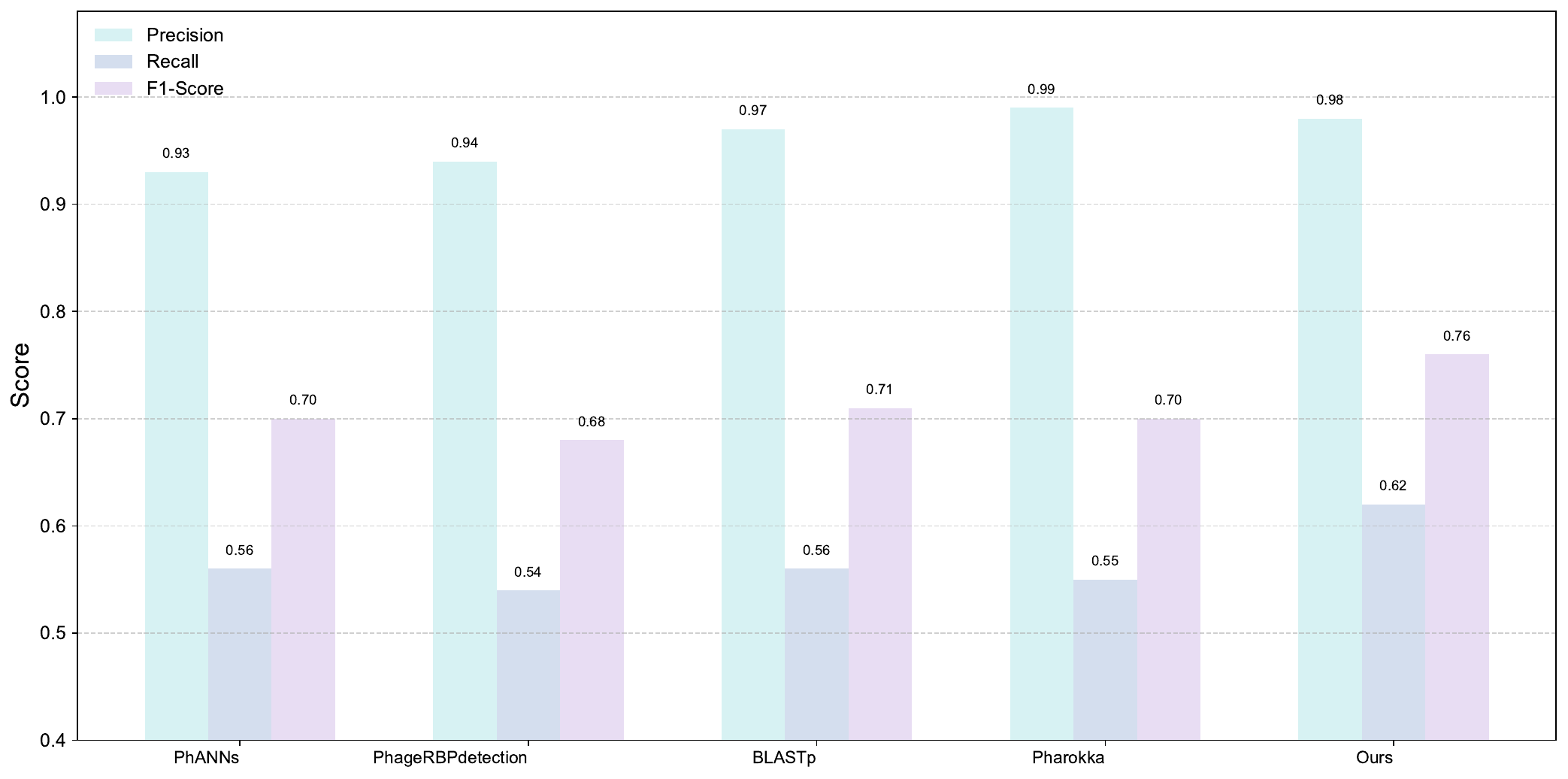}
    \vspace{-0.7cm}
    \caption{Bar chart evaluating the precision, recall and F1-Score of SeekRBP (denoted as "Ours") alongside existing machine learning-based tools (PhANNs, PhageRBPdetection) and homology-based methods (BLASTp, Pharokka).}
    \vspace{-0.5cm}
    \label{fig:sota}    
\end{figure}

\begin{table*}[!t]
\centering
\caption{Runtime comparison of competing methods on 1,000 protein sequences (unit: minutes). For SeekRBP, \textit{Sequence Only} uses only the sequence branch, \textit{Precomputed Structure} assumes that structures are already
available, and \textit{End-to-End} includes ColabFold structure prediction. Values in parentheses denote runtimes using public reference databases. All methods were run on an Intel\textsuperscript{\textregistered} Xeon\textsuperscript{\textregistered} Gold 6258R CPU with 40 cores and a Tesla A100 GPU when required.}
\label{tab:runtime_compare}

\small
\renewcommand{\arraystretch}{1.15}

\begin{tabular*}{\linewidth}
{@{\extracolsep{\fill}}lccccccc@{}}
\toprule
\multirow[c]{2}{*}{Method}
& \multicolumn{4}{c}{Competing methods}
& \multicolumn{3}{c}{SeekRBP workflows} \\
\cmidrule(lr){2-5}
\cmidrule(lr){6-8}

& PhANNs
& PhageRBPdetection
& BLASTp
& Pharokka
& Sequence Only
& Precomputed Structure
& End-to-End \\
\midrule

Runtime (min)
& 2.08
& 2.41
& 7.71 (51.86)
& 3.90 (14.28)
& 2.02
& 2.15
& 33.22 \\
\bottomrule
\end{tabular*}
\end{table*}

We further compared the precision, recall and F1-score of all competing methods (Fig. \ref{fig:sota}). The results revealed that SeekRBP yielded the best overall F1-score, which was approximately 7\% higher than the second best methods, while maintaining competitive precision and recall. Further error analysis indicated that most false positive sequences are labelled as structural proteins without detailed functions, which indicate that these ``incorrect'' predictions may be potential RBPs. Across five independent runs with different data partitions and model initializations, SeekRBP achieved a precision of $0.95 \pm 0.02$, a recall of $0.63 \pm 0.02$, and an F1-score of $0.76 \pm 0.01$ (Supplementary Table S1), demonstrating stable performance across random seeds. We also summarized model performance under thresholds ranging from 0.40 to 0.80 (Supplementary Table S2). Across this range, precision increased from 0.91 to 0.99, whereas recall decreased from 0.65 to 0.54 and the F1-score decreased from 0.76 to 0.70. These results illustrate the precision–recall trade-off associated with threshold adjustment: higher thresholds favor precision at the cost of reduced RBP recovery. Thus, users can select customized thresholds according to their actual application scenarios and performance preferences.

In addition, SeekRBP offers balanced performance and fast inference speed 
under sequence-only or precomputed-structure settings for large-scale RBP screening tasks. 
We benchmarked runtime across all tools on 1,000 protein sequences 
(Table \ref{tab:runtime_compare}). 
The Sequence Only and Precomputed Structure workflows consumed 2.02 and 2.15 minutes, respectively, whereas the End-to-End workflow, including ColabFold structure prediction, required 33.22 minutes. The Precomputed Structure workflow achieved computational efficiency comparable to PhANNs (2.08 min), while BLASTp and Pharokka required longer runtimes, particularly when using large public reference databases.

\subsection{Case study: application on \textit{Vibrio} phages}

To evaluate the biological relevance and generalizability of our proposed framework, we applied SeekRBP to an independent experimental dataset of \textit{Vibrio} phages. This dataset, generated by our collaborator Zhang et al. \citep{zhang2025isolation}, involved phage purification, culturing, and lytic induction. Then, the RBPs were identified by inspecting the structural domain using Hidden Markov Models, which is the same workflow as Pharokka. While an absolute ground truth for all RBPs is difficult to establish, we utilized this curated set (denoted as \textbf{Observed}) as our reference-based standard. In this analysis, we first used SeekRBP to identify putative RBPs and then compared their downstream utility for host prediction against the manually curated set.

\begin{figure}[htbp]
    \centering
    \includegraphics[width=1.0\linewidth]{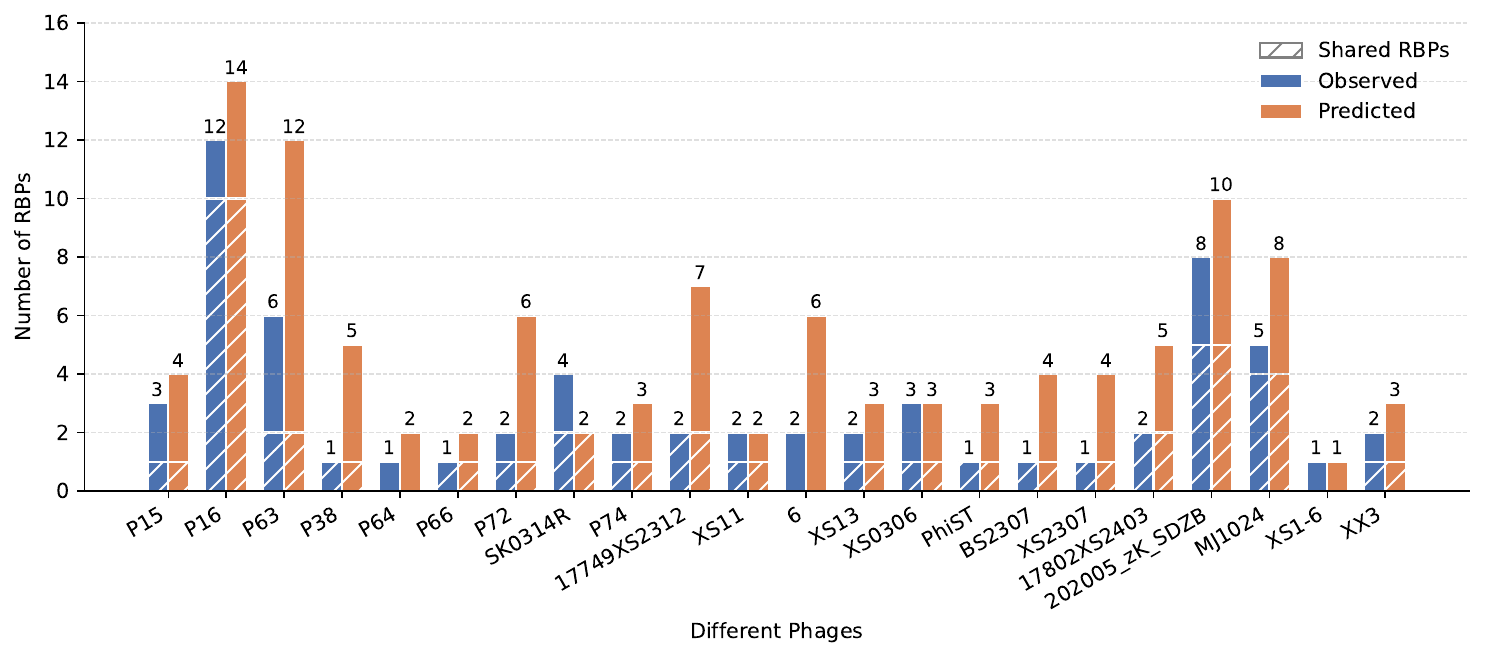}
    \vspace{-0.5cm}
    \caption{Comparison of the number of predicted RBPs and manually observed RBPs across individual \textit{Vibrio} phages. The x-axis denotes different \textit{Vibrio} phages, and the y-axis represents the number of RBPs. Our model identified more candidate RBPs than those detected manually.}
    \vspace{-0.5cm}
    \label{fig:vibrio1}
\end{figure}

As shown in Fig. \ref{fig:vibrio1}, SeekRBP identifies a broader set of RBPs than the manually curated reference, while still recovering a substantial fraction of the known RBPs across various phage genomes. Beyond count-based comparisons, we conducted a structure-based analysis using TM-scores to evaluate the structural plausibility of the newly predicted RBPs. Specifically, we compared the TM-scores of manually selected RBPs missed by our framework (the \textbf{Observed} set) against those identified by SeekRBP but absent from manual annotations (the \textbf{Predicted} set). We additionally included a control set of annotated non-RBP proteins in the \textit{Vibrio} dataset.
A TM-score above 0.5 generally indicates that two proteins share the same global fold. As illustrated in Supplementary Fig. S5, while the predicted RBPs exhibit a wider variance in TM-scores, the majority exceed the 0.5 threshold. This indicates that SeekRBP successfully detects divergent RBP candidates with valid structural folds, potentially capturing functional proteins that were overlooked during manual curation.

\begin{figure}[htbp]
    \centering
    \vspace{-0.5cm}
    \includegraphics[width=1.0\linewidth]{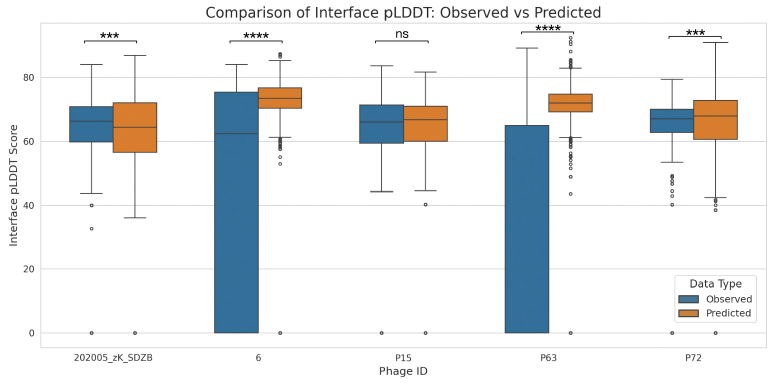}
    \vspace{-0.5cm}
    \caption{Docking quality (interface pLDDT) between RBPs and outer membrane proteins from host genomes estimated by AlphaFold3. The \textbf{Predicted} set exhibits higher interface pLDDT scores than the \textbf{Observed} set in some phages, reflecting phage-dependent differences.}
    \vspace{-0.5cm}
    \label{fig:vibrio3}
\end{figure}

\vspace{-0.5cm}

Furthermore, we investigated whether these newly predicted RBPs could enhance downstream host prediction. We compared the utility of the \textbf{Observed} and Predicted sets using two complementary validation approaches. First, we evaluated their performance in a learning-based phage-host interaction prediction task. Using a standard multilayer perceptron classifier within a 10-fold stratified cross-validation framework (Supplementary S1.7), the \textbf{Predicted} set yielded a higher mean AUROC (0.7367) compared to the \textbf{Observed} set (0.7132). Additionally, the standard deviation decreased from 0.099 to 0.073, indicating improved predictive stability (Supplementary Fig. S6). Second, as a complementary computational assessment, we employed AlphaFold3 to estimate the predicted interface confidence, measured by interface pLDDT, between the RBP candidates and host outer membrane proteins (Supplementary S1.8). Phage-specific statistical comparisons revealed significant differences between the \textbf{Observed} and \textbf{Predicted} groups for four of the five examined phages, whereas no significant difference was detected for P15. However, the magnitude and direction of these differences varied across phages. Thus, Fig.~\ref{fig:vibrio3} indicates phage-dependent differences in predicted interface confidence rather than a consistent overall improvement in binding stability. Because interface pLDDT reflects confidence in the predicted interface rather than experimentally measured binding stability, this analysis is presented as complementary computational evidence rather than direct functional validation. Together, these results suggest that the additional RBP candidates identified by SeekRBP may supplement sparse reference annotations.

\section{Discussion}

In this study, we proposed SeekRBP, an integrative framework for phage RBP identification that combines sequence- and structure-derived representations with a reinforcement learning-inspired adaptive negative sampling strategy. By formulating negative sample selection as a sequential decision process guided by training feedback, SeekRBP effectively addresses the extreme class imbalance inherent in RBP datasets. Our bandit-based approach offers a computationally efficient approximation to full reinforcement learning; it enables a dynamic emphasis on hard, informative negatives without requiring explicit state modeling, and pairs effectively with the multimodal sequence–structure fusion module. Comprehensive evaluations demonstrate that SeekRBP improves recall and overall predictive performance compared to existing methods. Notably, the framework successfully identified unannotated RBPs in \textit{Vibrio} phages, providing valuable biological context that enhances downstream host prediction.

While SeekRBP provides a robust computational tool for phage RBP identification, several avenues remain for future optimization. First, we plan to extend the framework to support receptor-specific and multi-class prediction tasks. This expansion will involve incorporating richer structural contexts, particularly protein–receptor interaction dynamics, to better capture binding specificity. Second, considering the high GPU demands of processing 3D structures, we introduce a CPU-only mode on GitHub. Furthermore, we leverage feature compression to rapidly generate 3D structure embeddings as an efficient alternative to AlphaFold3. Finally, as additional experimental annotations become available, iterative retraining will be essential to strengthen model generalization, supporting practical applications in phage therapy and antimicrobial discovery.

%%%%%%%%%%%%%%%%%%%%%%%%%%%%%%%%%%%%%%%%%%%%%%%%%%%%%%%%%%%%%%%%%%%%%%%%%%%%%%%%%%%%%
%
%     please remove the " % " symbol from \centerline{\includegraphics{fig01.eps}}
%     as it may ignore the figures.
%
%%%%%%%%%%%%%%%%%%%%%%%%%%%%%%%%%%%%%%%%%%%%%%%%%%%%%%%%%%%%%%%%%%%%%%%%%%%%%%%%%%%%%%

%\section*{Acknowledgements}
\vspace{-0.3cm}
\subsection*{Data availability}
All data and codes used for this study are available online via: \\
\href{https://github.com/Saillxl/SeekRBP}{https://github.com/Saillxl/SeekRBP}.

\vspace{-0.3cm}
\subsection*{Funding}
This work was supported by the RGC GRF CityU 11209823, City University of Hong Kong projects 9667256, 9678241, 7020092,  Institute of Digital Medicine, and Guangdong Provincial Key Laboratory of Wastewater Information Analysis and Early Warning 217310019.

\vspace{-0.7cm}
\tiny
\bibliographystyle{natbib}
\bibliography{Document}

\end{document}